\documentclass[]{article}

\usepackage[utf8]{inputenc}
\usepackage{cite}
\usepackage{amssymb}
\usepackage[shortlabels]{enumitem}
\usepackage{graphicx} 
\usepackage{multicol}
\usepackage{epstopdf}
\usepackage{booktabs}
\usepackage{amsmath}
\usepackage{multirow,hhline}
\usepackage{spconf}

\usepackage[hyphens]{url}

\twocolumn
\def\ninept{\def\baselinestretch{.95}\let\normalsize\small\normalsize}

\date{June 2021}

\title{ON THE INVERTIBILITY OF A VOICE PRIVACY SYSTEM\\ USING EMBEDDING ALIGNMENT}

\name{Pierre Champion$^{2, 3, \star}$\quad
Thomas Thebaud$^{1, 2, \star}$\quad
Gaël Le Lan$^{1}$\quad
Anthony Larcher$^{2}$\quad
Denis Jouvet$^{3}$\thanks{This work was supported in part by the French National Research Agency under project DEEP-PRIVACY (ANR-18-CE23-0018) and Région Grand Est. Experiments were carried out using the Grid’5000 testbed, supported by a scientific interest group hosted by Inria and including CNRS, RENATER and several Universities as well as other organizations.}
}

\address{$^{1}$ Orange, France \\
$^{2}$ LIUM - EA4023, Le Mans Université, Avenue Olivier Messiaen,72085 LE MANS CEDEX 9, France \\
$^{3}$ Université de Lorraine, CNRS, Inria, LORIA, F-54000 Nancy, France\\
$^{\star}$ equal contribution from authors
}

\begin{document}
%
\maketitle

\begin{abstract}
    This paper explores various attack scenarios on a voice anonymization system using embeddings alignment techniques. 
    We use Wasserstein-Procrustes (an algorithm initially designed for unsupervised translation) or Procrustes analysis to match two sets of $x$-vectors, before and after voice anonymization, to mimic this transformation as a rotation function. 
    We compute the optimal rotation and compare the results of this approximation to the official Voice Privacy Challenge results. 
    We show that a complex system like the baseline of the Voice Privacy Challenge can be approximated by a rotation, estimated using a limited set of $x$-vectors. 
    This paper studies the space of solutions for voice anonymization within the specific scope of rotations. 
    Rotations being reversible, the proposed method can recover up to 62\% of the speaker identities from anonymized embeddings.


\end{abstract}

\section{Introduction}
\label{sec:introduction}
Modern supervised deep learning algorithms require a large amount of data to be trained. 
To address this, service providers collect, process, and store personal data in centralized servers, raising serious concerns regarding their customer's data privacy.
Recent regulations, e.g., the General Data Protection Regulation (GDPR)~\cite{gdpr} in the European Union, 
emphasize the need for service providers to ensure privacy preservation and protection of personal data. 
As speech data can reflect both biological and behavioral characteristics of the speaker, it is qualified as personal data \cite{nautschGDPRSpeechData2019}.

The ISO/IEC international Standard 24745 on biometric data protection~\cite{iso24754} defines \textbf{unlinkability} and \textbf{non-invertibility} criteria for privacy protection as follows:
\begin{itemize}
    \item \textbf{Unlinkability} means that anonymized data processed in a privacy-relevant manner should not be linkable to any other set of data (anonymized or not) outside of the domain. Protected data processed in the same privacy-relevant manner must be discriminative enough to satisfy the service provider requirements, but not attackers.
    \item \textbf{Non-invertibility} means that it should be computationally infeasible\footnote{Cannot be solved using an algorithm with polynomial complexity.} to obtain the clear data that led to any given anonymized data.
\end{itemize}

To achieve \textbf{unlinkability}, speaker anonymization \cite{fangSpeakerAnonymizationUsing2019,magarinos2017reversible} is performed to suppress the personally identifiable para-linguistic information from a speech utterance while maintaining the linguistic content.
Recently, Fang et al.~\cite{fangSpeakerAnonymizationUsing2019} proposed a speaker anonymization system based on the $x$-vector paradigm and a voice conversion method.
This system was used as a baseline in the first edition of the Voice privacy Challenge (VPC). 
The quality of anonymization is assessed using a state-of-the-art speaker verification system, which evaluates the \textbf{unlinkability} criteria defined in ISO/IEC 24745.

In this work, we propose to invert the VPC baseline's anonymization system using embedding alignment algorithms.
In a first step, we follow the scenarios of the VPC in terms of attacker knowledge about the anonymization system~\cite{EvaluatingVoiceConversionbased2019,tomashenkoVoicePrivacy2020Challenge}.
The challenge assumes that the attacker has a set of clear speaker $x$-vectors with the corresponding anonymized $x$-vectors.
Having this mapping allows us to approximate the anonymization function with a rotation, using a supervised Procrustes Analysis~\cite{gower1975generalized}.
We propose a more restrictive scenario where the attacker does not know which clear $x$-vector corresponds to the anonymized $x$-vector.
In this scenario, we use an unsupervised embedding alignment algorithm named Wasserstein-Procrustes~\cite{grave2018unsupervised}.

Once the anonymized $x$-vector is projected to estimate his corresponding clear $x$-vector, we evaluate the \textbf{linkability} performance between speech accessible to the attacker (enrollment) and speech anonymized by the service provider (trials). \textbf{Invertibility} is evaluated by measuring how well an attacker can invert the anonymized $x$-vectors of a service provider (trials).
%
The main contributions of this paper are: (i) the approximation of a speaker anonymization system by a rotation, (ii) the use of supervised and unsupervised embedding alignment to estimate this rotation, (iii) the first (to our knowledge)  \textbf{invertibility} attack of a speaker anonymization system.
In Section \ref{sec:voice_privacy}, we describe the VPC goals, baseline systems, data, and scenarios. 
Section \ref{sec:unsupervised} presents supervised and unsupervised alignment techniques used to perform the attack.
Section \ref{sec:scenarios} introduces the attack scenarios. 
Experimental protocol and results are detailed 
in Section \ref{sec:experiments} and Section \ref{sec:conclusion} discusses the outcomes of this work and puts them in perspective for future research.

\section{Voice Privacy challenge}
\label{sec:voice_privacy}
The VPC 2020~\cite{tomashenkoVoicePrivacy2020Challenge} proposed an evaluation framework, dataset, and attack scenarios, which are presented in this section, to guide and facilitate the development of privacy-preserving approaches in the speech domain.

\subsection{The speaker anonymization system}

\begin{figure}[ht]
  \centering
  \includegraphics[width=1.00\linewidth]{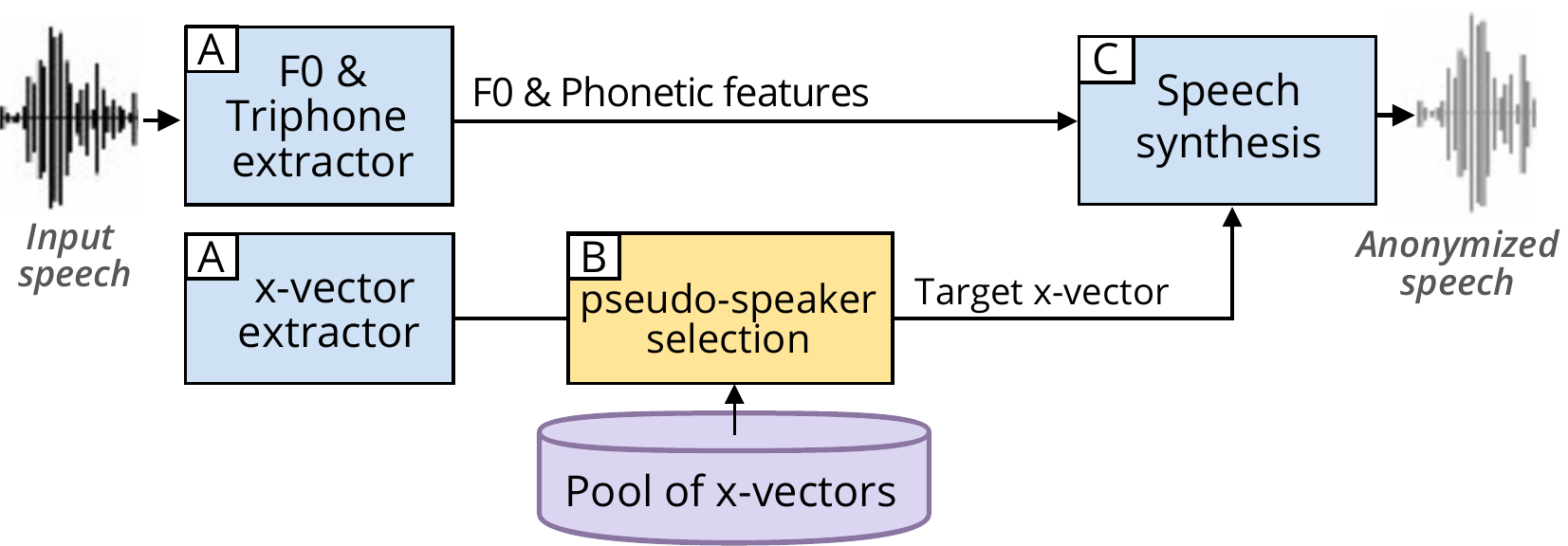}
  \caption{The Voice Privacy speaker anonymization pipeline.
  }
  \label{fig:baseline.png}
\end{figure}

The speaker anonymization system used in this work anonymizes speech segments using a $x$-vector-based approach~\cite{fangSpeakerAnonymizationUsing2019}.
Speaker identity and linguistic content are first extracted from an input speech utterance.
Assuming that those features are disentangled, an anonymized speech waveform is generated by altering only the features that encode the speaker's identity.
The anonymization system depicted in {Figure \ref{fig:baseline.png}} can be decomposed into three groups of modules.
Modules from the \textit{group A} extract different features from the source signal: the fundamental frequency, the phonetic features encoding articulation of speech sounds and the speaker's $x$-vector.
\textit{The module B} derives a new target identity. The $x$-vector from each source input speaker is compared to a pool of external $x$-vectors in order to select the 200 furthest vectors; 100 of them are randomly selected and averaged to create an anonymized target $x$-vector identity. 
Finally, \textit{the module C} synthesizes a speech waveform from the target $x$-vector together with the original phonetic features and F0.
Speaker anonymization is achieved by selecting a private target $x$-vector.

\subsection{Dataset}
\label{sec:data}
In the VPC, the evaluation dataset is built from LibriSpeech \textit{test-clean}~\cite{Librispeech}.
Details about the number of speakers and utterances in the enrollment and trial datasets are reported in Table \ref{tab:data}.
The speech segments from 40 speakers are used to create two sets: an $\mathbf{Enroll}$ and a $\mathbf{Trials}$ set, both containing similar Female/Male ratios.
Speakers from the $\mathbf{Enroll}$ set are all contained in the $\mathbf{Trials}$ sets, but their utterances are distinct between sets.
\begin{table}[ht]
    \centering
    \caption{Statistics of the evaluation dataset. F and M indices refer to Female and Male speakers respectively.}
    \begin{tabular}{c c c c}
        \toprule
        Set & Speakers & Utterances & Gender \\
        \midrule
        $\mathbf{Enroll}$ & 29 & 438 & Both\\
        $\mathbf{Enroll_F}$ & 16 & 254 & Female \\
        $\mathbf{Enroll_M}$ & 13 & 184 & Male\\
        \midrule
        $\mathbf{Trials}$ & 40 & 1496 & Both\\
        $\mathbf{Trials_F}$ & 20 & 734 & Female\\
        $\mathbf{Trials_M}$ & 20 & 762 & Male\\
        \bottomrule
    \end{tabular}
    \label{tab:data}
\end{table}


\subsection{VPC attack scenarios}
\label{subsec:VPC_scenars}

Speech data anonymized by the service provider is referred to as the trial dataset, and clear speech accessible to the attacker is referred to as the enrollment dataset.
During the challenge, three sets of tests are performed, following the \textit{Ignorant}, \textit{Lazy-Informed}, and \textit{Semi-Informed} attacker scenarios~\cite{EvaluatingVoiceConversionbased2019,brij_journal}. 
In the \textit{Ignorant} scenario, the attacker is unaware that speech was transformed. 
Thus, he performs a linkability test between the anonymized trial (denoted as $\mathbf{Trial_{Anon}}$) and the clear, non-anonymized enrollment dataset (denoted as $\mathbf{Enroll}$) using 
an automatic speaker verification (ASV) system with an $x$-vector extractor~\cite{snyder2018xvector} trained on the clear speech of LibriSpeech \textit{train-100}~\cite{Librispeech,tomashenkoVoicePrivacy2020Challenge}.

In the \textit{Lazy-Informed} and \textit{Semi-Informed} scenarios, the attacker has partial knowledge of the system and is able to anonymize the enrollment utterances using the same anonymization system but not the same target $x$-vector.
The target $x$-vector chosen for each enrollment speaker differs from the target $x$-vector chosen for the trial speakers as the attacker does not know the randomly selected $x$-vectors used to generate the target identity. 
The difference between \textit{Lazy-Informed} and \textit{Semi-Informed} lies in the data used to train the $x$-vector extractor, for the \textit{Lazy-Informed} scenario the $x$-vector extractor is trained on clear, non-anonymized speech.

The most powerful attacker defined by the challenge is the \textit{Semi-Informed} one. He has the same knowledge of the anonymization system as the \textit{Lazy-Informed}, he is also able to anonymize the enrollment utterances, and additionally, he anonymizes the clear LibriSpeech \textit{train-100} dataset to retrain the $x$-vector extractor on anonymized data.
Being computed by a retrained $x$-vector extractor, the $x$-vectors datasets used by the \textit{Semi-Informed} attacker give better results.

In this paper, we provide results that follow the \textit{Lazy-Informed} and \textit{Semi-Informed} attacker scenarios of the VPC, but also propose additional scenarios that explore more and less constraining hypotheses for respectively unsupervised and oracle attackers.

\section{Supervised and unsupervised alignment algorithms}
\label{sec:unsupervised}
Computing the alignment of two embeddings of high dimensional real vectors is a fundamental problem in machine learning, with applications for unsupervised word and sentence translation~\cite{biswas2020aligning, grave2018unsupervised,rapp1995identifying,fung1995compiling,bojanowski2017unsupervised}.

\subsection{Procrustes Analysis}
\label{subsec:algo_procrustes}
Let $\mathbf{A}$ and $\mathbf{B}$ be two sets of $N$ high dimensional real vectors of dimension $d$.
We want to find the optimal rotation $\mathbf{W}\in \mathbb{R}^{d\times d}$ that minimize the squared distance between both sets:
\begin{equation}
\label{eq:dist1}
    \min_{\mathbf{W}\in \mathbb{R}^{d\times d}} ||\mathbf{A}\mathbf{W} - \mathbf{B} ||^2_2
\end{equation}
For correctly matched sets $\mathbf{A}$ and $\mathbf{B}$ (the $n^{th}$ element of $\mathbf{A}$ corresponds to the $n^{th}$ element $\mathbf{B}$, $\forall n \in [\![1,N]\!]$), we can directly use Procrustes analysis~\cite{gower1975generalized} to compute optimal $\mathbf{W}$.

This approach is well suited for \textbf{supervised} scenarios since it requires access to the labels of both sets. For two unlabeled sets, an unsupervised alignment algorithm is required.


\subsection{Wasserstein-Procrustes}
\label{subsec:algo_Wasserstein}

Grave et al.~\cite{grave2018unsupervised} proposed an unsupervised algorithm to align sets of language-dependent word embeddings to perform unsupervised translation. 
The authors proposed a stochastic optimization, switching between minimizing the Wasserstein~\cite{ruschendorf1985wasserstein} distance between sets and finding the optimal rotation using the Procrustes analysis~\cite{gower1975generalized}, to find the rotation that optimally lowers the distance between the two sets of embeddings, as well as their one-to-one mapping. 
In the rest of the paper, we will use this algorithm to align speaker embedding sets in \textbf{unsupervised} scenarios.

\section{Invertibility Attack scenarios}
\label{sec:scenarios}
This paper explores invertibility attack scenarios (and their variations) that depend on different dataset accessibility hypotheses.

\subsection{Dataset accessibility hypotheses}
\begin{figure}[ht]
    \centering
    \includegraphics[width=0.99\linewidth]{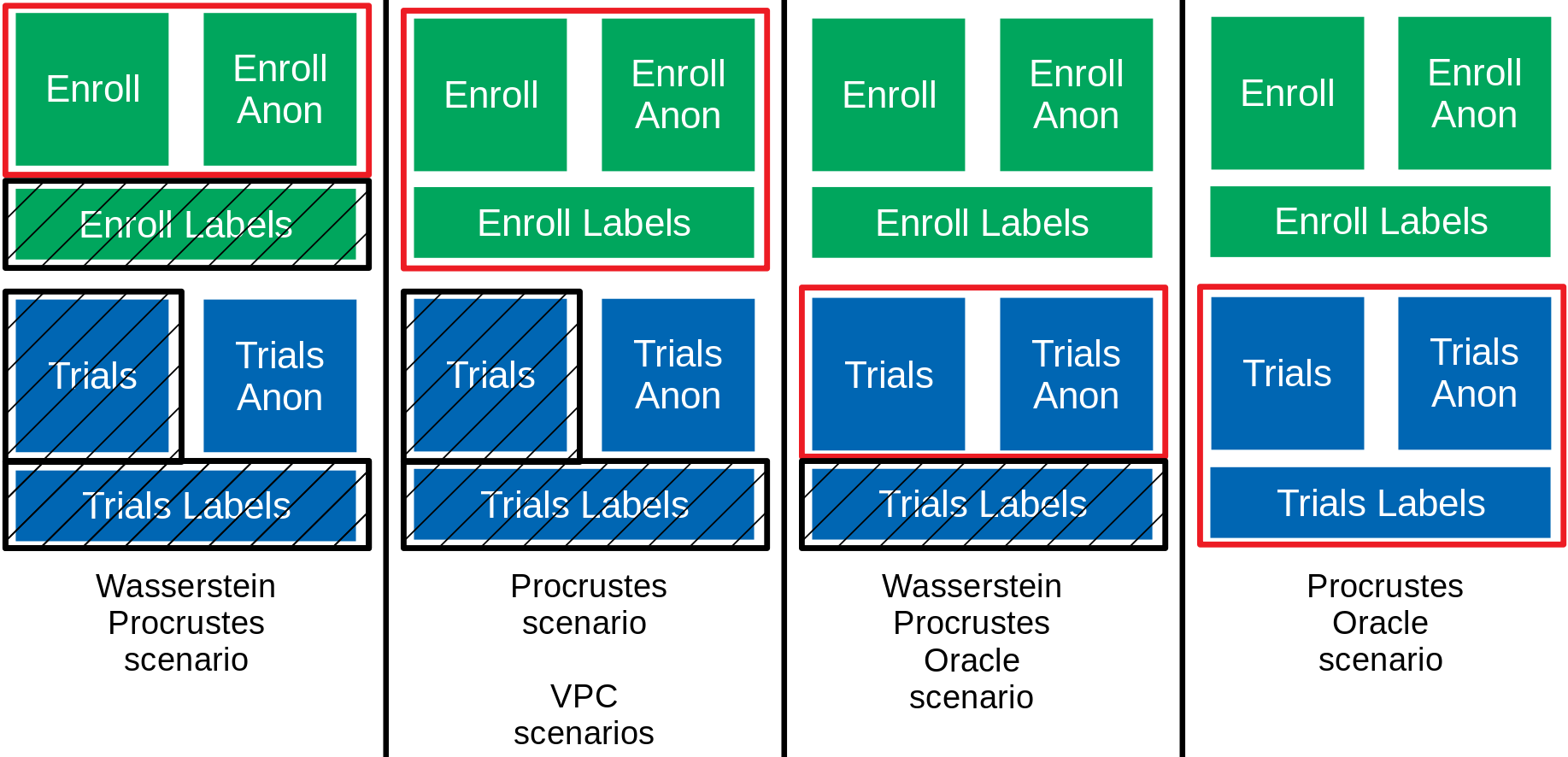}
    \caption{Schematic representation of the sets used for different scenarios. Figure best viewed in color.}
    \label{fig:scenars}
\end{figure}

The datasets accessibility hypotheses are summarized in Figure \ref{fig:scenars} for the different scenarios detailed in the Section \ref{subsec:scenario}.
Red boxes show data available to the attacker in a given hypothesis.  
Black hatched boxes show data inaccessible to the attacker in a given hypothesis.
We call supervised the scenarios where labels are available and unsupervised the ones where labels are inaccessible.
The scenarios where the attacker has access to the clear $\mathbf{Trials}$ are not realistic, but they allow to test the rotation effectiveness in the worst condition for the user.
Regardless of the available data, the performances are evaluated using the $\mathbf{Trials}$, $\mathbf{Trials_{Anon}}$ and $\mathbf{Enroll}$ sets.

\subsection{Scenarios}
\label{subsec:scenario}
We explore invertibility attacks on the speaker anony\-mization system in supervised and unsupervised conditions (described in sections \ref{subsubsec:scenar_supervised} and \ref{subsubsec:scenar_unsupervised} respectively).
We approximate the anonymization function in the $x$-vector domain by a rotation that is estimated by a supervised (Procrustes, \ref{subsec:algo_procrustes}) or unsupervised algorithm (Wasserstein-Procrustes, \ref{subsec:algo_Wasserstein}).
Those two invertibility attacks are realistic, but we also compare them for reference to less realistic \textit{Oracle} versions, where the attacker has also access to clear $\mathbf{Trials}$ (see Section \ref{subsubsec:scenar_oracle}).

\begin{figure}[ht]
    \centering
    \includegraphics[scale=0.185]{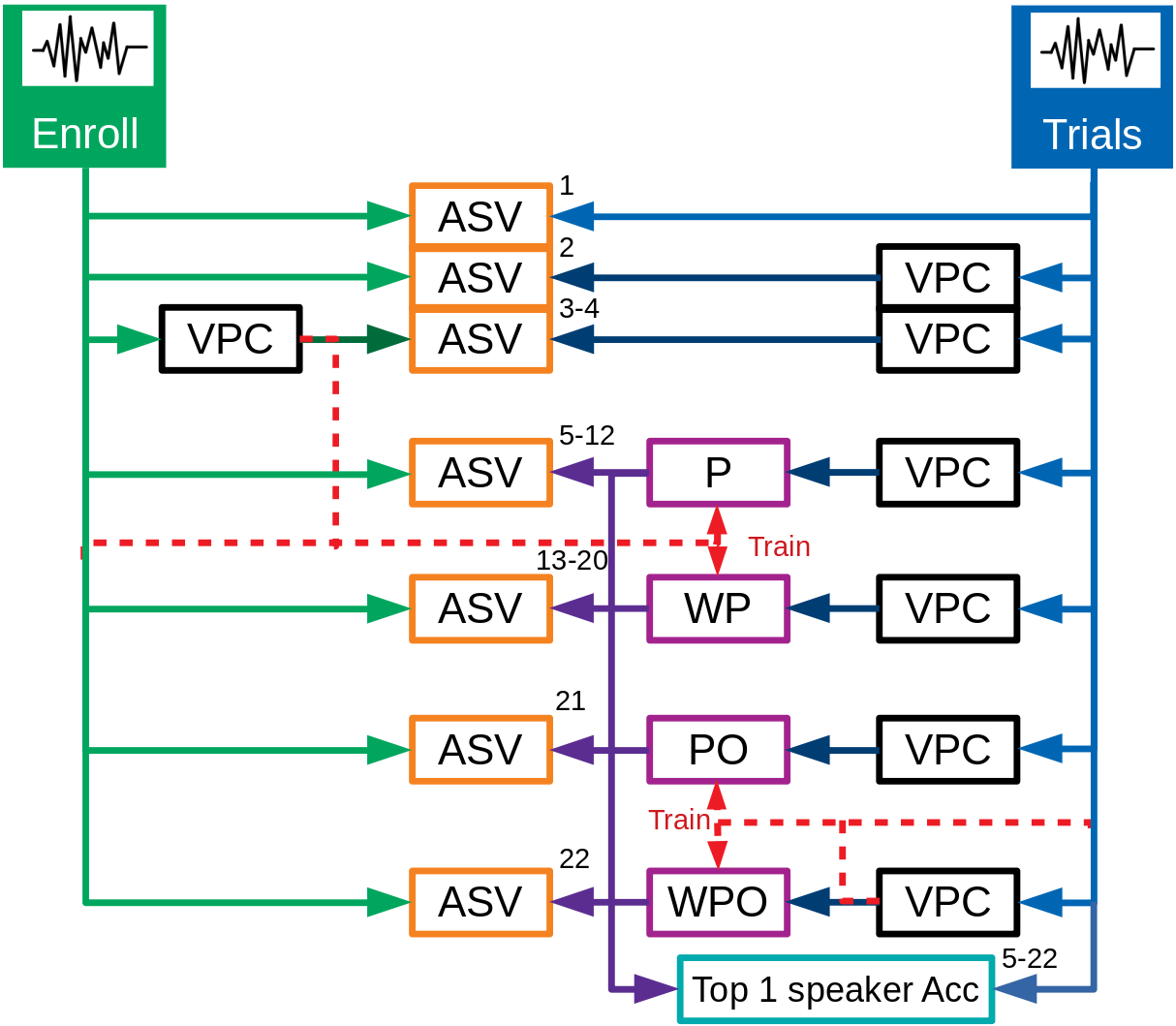}
    \caption{Schematic representation of the different attacks. Figure best viewed in color.}
    \label{fig:attacks}
\end{figure}

Figure \ref{fig:attacks} (better seen in color) presents a schematic representation of the different considered attacks and or their evaluation.
The $\mathbf{Enroll}$ sets are presented in green, the $\mathbf{Trials}$ sets in blue. 
The red arrows represent the datasets used to estimate the rotations (purple blocks). 
The results computed by the orange (automatic speaker verification \textit{ASV}) and cyan (Top1 speaker acc.) blocks are reported in their respective lines in Table \ref{tab:results}: the numbers next to the ASV boxes refer to the lines of Table \ref{tab:results} that present the corresponding results.
$\mathbf{Enroll}$ and $\mathbf{Trials}$ blocks are sets of waveforms, VPC blocks refer to the $x$-vector-based speaker anonymization system, which takes a waveform as input, and outputs a waveform corresponding to the anonymized utterance. 
Purple blocks align sets of embeddings following different scenarios, which are described in this section: Procrustes (P); Wasserstein-Procrustes (WP); Procrustes oracle (PO); or Wasserstein-Procrustes oracle (WP0).
Note that P, WP, PO, and WPO handle $x$-vectors, and that the speaker verification also relies on $x$-vectors. However, for better readability, the computation of the $x$-vectors is not explicitly shown in the figure.
As an example, to obtain the results of the $12^{th}$ line of Table \ref{tab:results}: the $\mathbf{P}$ matrix is computed using the $\mathbf{Enroll}$ and $\mathbf{Enroll_{Anon}}$ sets, then applied on the $\mathbf{Trials_{Anon}}$ set, the EER is computed after scoring the $\mathbf{Enroll}$ set against the projected $\mathbf{P^T}\times\mathbf{Trials_{Anon}}$ with the orange ASV block, and finally the Top 1 speaker accuracy is computed between $\mathbf{Trials}$ and $\mathbf{P^T}\times\mathbf{Trials_{Anon}}$. 

For all following scenarios, once the rotation matrix $\mathbf{W}\in\mathbb{R}^{d\times d}$ is estimated, the set of $\mathbf{Trials_{Anon}}$ $x$-vectors is inverted using the transposed $\mathbf{W}$:
\begin{equation}
    \label{eq:project}
    \mathbf{Trials^*_{W}} = \mathbf{W^T}\times \mathbf{Trials_{Anon}}
\end{equation}

\subsubsection{Supervised scenario: Procrustes}
\label{subsubsec:scenar_supervised}
Our first scenario, the Procrustes attack, follows the rules of the VPC challenge (the hypotheses described in Section \ref{sec:voice_privacy} are the same).
It uses a rotation $\mathbf{P}$, computed in a supervised manner.
We apply Procrustes on the $\mathbf{Enroll}$ and $\mathbf{Enroll_{Anon}}$ sets, knowing the one-to-one correspondence between them (thanks to the $\mathbf{Enroll_{labels}}$ knowledge). 
\begin{equation}
\small
    \mathbf{P} = Procrustes(\mathbf{Enroll}, \mathbf{Enroll_{Anon}}, \mathbf{Enroll_{labels}})
\end{equation}
Then $\mathbf{Trials_{Anon}}$ are inverted using equation \ref{eq:project} with $\mathbf{W}=\mathbf{P}$

The goals of this first experiment are to measure:
\begin{itemize}
    \item How well a rotation can approximate the VPC system in the $x$-vector domain by comparing the EER to the ones obtained in similar conditions for the different scenarios of the VPC.
    \item How many $\mathbf{Trials_{Anon}}$ $x$-vectors can be inverted well enough to recognize their source speaker, using the reversed rotation. 
\end{itemize}

\subsubsection{Unsupervised scenario: Wasserstein-Procrustes}
\label{subsubsec:scenar_unsupervised}
This second experiment explores the performance of an unsupervised algorithm for the invertibility attack.
The hypothesis in this scenario presents a slight difference with the VPC ones: the attacker does not have access to the labels $\mathbf{Enroll_{Labels}}$, hence the use of the Wasserstein-Procrustes algorithm.

The optimal rotation $\mathbf{WP}$ can be computed using the following equation:
\begin{equation}
\small
    \mathbf{WP}= Wasserstein\_Procrustes(\mathbf{Enroll}, \mathbf{Enroll_{Anon}})
\end{equation}

The goal of this scenario is to evaluate the degradation of performance when not using $\mathbf{Enroll_{Labels}}$.
Due to the VPC anonymization process, some $x$-vectors of the $\mathbf{Enroll}$ and $\mathbf{Enroll_{Anon}}$ sets could be mismatched (misaligned) during the unsupervised training.
Every mismatch error will contribute to degrade the alignment and lower the training and testing performances.

We also apply the variations presented in Section \ref{subsec:variations} to this experiment.
The results are presented in lines 13 to 20 of Table \ref{tab:results}.

\subsubsection{Oracle scenarios}
\label{subsubsec:scenar_oracle}
This third experiment probes the optimal performances one can get while approximating a speech anonymization system by a supervised or unsupervised rotation estimated in the $x$-vector domain.

We suppose that an oracle has access to the $\mathbf{Trials}$, $\mathbf{Trials_{Anon}}$ and $\mathbf{Trials_{Labels}}$ sets, meaning it can compute the best approximation possible on the evaluation data.
The rotation matrix $\mathbf{PO}$ is computed using Procrustes:
\begin{equation}
\small
    \mathbf{PO} = Procrustes(\mathbf{Trials}, \mathbf{Trials_{Anon}}, \mathbf{Trials_{labels}})
\end{equation}
And the rotation matrix $\mathbf{WPO}$ is computed using Wasserstein-Procrustes:
\begin{equation}
\small
    \mathbf{WPO} = Wasserstein\_Procrustes(\mathbf{Trials}, \mathbf{Trials_{Anon}})
\end{equation}
So the $\mathbf{Trials_{Anon}}$ can be inverted with the obtained rotation matrices (same process as in equation \ref{eq:project}).

For the oracle scenarios, the rotation is directly computed on the evaluation data. This gives the performance upper bound for the first two experiments (lines 12 and 20 of Table \ref{tab:results}).
Results are presented in lines 21 and 22 of Table \ref{tab:results}.

{\let\thefootnote\relax\footnote{{The python code of the experiments is available at \url{https://github.com/deep-privacy/x-vector-procrustes}}}}

\subsection{Experimental variations}
\label{subsec:variations}
\subsubsection{Principal component analysis}
\label{subsec:PCA}
To improve the attack performance, we extend the range of our experiments to modified $x$-vectors domains.
We apply a dimensional reduction technique to the $x$-vectors sets: principal component analysis~\cite{jolliffe2005principal} (PCA). 
Reducing the number of dimensions reduces the candidate rotations manifold, simplifying the search for the optimal one. 
The PCA also orders the dimensions precisely: the dimensions with the higher variance represented are placed first. 
This means that applying PCA on two vectors sets acts as a pre-alignment, easing the following alignment process.

For every experiment scenario using the PCA variation, we used a reduction in 70 dimensions (originally 512 for the $x$-vectors), and the total explained variance ratio was always above 98.0\%.


\subsubsection{Gender dependent training}
\label{subsec:gender}
As defined by the VPC evaluation rules, the \textbf{unlinkability} performances on Female and Male speakers are measured separately.
A gender-dependent variation trains two separated rotations to improve the attack performance, one only on Female sets and the other only on Male sets.

\subsubsection{Retrained original $x$-vector extractor}
\label{subsubsec:retrained}
Corresponding to the \textit{Lazy-Informed} and \textit{Semi-Informed} attacker of VPC, the $x$-vector extractor is either trained on clear, original speech or on anonymized speech.

\section{Experiments}
\label{sec:experiments}

\begin{table*}[t]
    \caption{Experimental results for the considered attack scenarios. Lines 1-4 correspond to the baseline attack scenarios of the VPC. Lines 5-12, 13-20 and 21-22 correspond to our rotation-based attack scenarios described respectively in the sections \textit{Supervised scenario} (\ref{subsubsec:scenar_supervised}), \textit{Unsupervised scenario} (\ref{subsubsec:scenar_unsupervised}) and \textit{Oracle scenarios} (\ref{subsubsec:scenar_oracle}). The variations corresponding to the columns \emph{Gender dependent}, \emph{PCA} and \emph{$x$-vector extractor} are described in the section \textit{Experimental variations} (\ref{subsec:variations}).}
    \centering
    \begin{tabular}{c c c c c c c c c}
    \toprule
         \multicolumn{2}{c}{} & Gender & \multirow{2}{*}{PCA} & $x$-vector & \multicolumn{2}{c}{EER} & \multicolumn{2}{c}{Top 1 speaker Acc.} \\
         \multicolumn{2}{c}{} & dependent  & & extractor & F & M & F & M \\
    \midrule
         1 & Baseline (clear data)      &                &           & original  & 10.3\%   & 2.9\%    &           &       \\
         2 & VPC - \textit{Ignorant}   &                &           & original  & 49.0\%   & 42.6\%   &           &       \\
         3 & VPC - \textit{Lazy-Informed}   &                &           & original  & \textbf{29.4}\%   & \textbf{29.1}\%   &           &       \\
         4 & VPC - \textit{Semi-Informed}   &                &           & retrained      & \textbf{17.1}\%   & \textbf{14.1}\%   &           &       \\
    \midrule
         5 &            &               &           & original  & 41.9\%   & 30.6\%   & 25.5\%   & 36.6\% \\
         6 &            & \checkmark    &           & original  & 40.1\%   & 31.0\%   & 26.6\%   & 45.8\% \\
         7 &            &               & \checkmark & original & 32.6\%   & 32.7\%   & 27.1\%   & 40.8\% \\
        \vspace{0.0cm}
         8 & \multirow{2}{*}{Procrustes}           & \checkmark   & \checkmark & original & \textbf{25.4}\%   & \textbf{24.4}\%   & \textbf{30.5}\%   & \textbf{50.0}\% \\
         \cline{3-9}
         
         \noalign{\vskip 0.1cm} 9 &  & &           & retrained      & 29.0\%   & 23.6\%   & 58.7\%   & 59.2\% \\
         10 &            & \checkmark    &           &  retrained     & 27.0\%   & 21.1\%   & 54.8\%   & 56.6\% \\
         11 &           &              & \checkmark & retrained     & 21.5\%   &  23.1\%   & 51.9\%   & 57.0\% \\
         12 &            & \checkmark   & \checkmark &  retrained    & \textbf{14.6}\%   &  \textbf{13.1}\%   & \textbf{59.8}\%   & \textbf{60.0}\% \\
    \midrule
         13 &            &              &           & original  &  43.6\%   & 33.4\%   &  25.2\%   &  22.1\% \\
         14 &           & \checkmark   &           & original  &   40.7\%   & 35.9\%   &  26.6\%   &  20.9\% \\
         15 &           &               & \checkmark & original &  36.3\%   & 35.4\%   &  \textbf{24.1}\%   &  38.6\% \\
         \vspace{0.0cm}
         16 & Wasserstein & \checkmark    & \checkmark & original &  \textbf{26.4}\%   & \textbf{25.2}\%   & \textbf{24.1}\%   &  \textbf{39.4}\% \\
         \cline{3-9}
         \noalign{\vskip 0.1cm} 17 & Procrustes &             &           & retrained      & 31.4\%   & 24.2\%   & 57.2\%   & 60.2\% \\
         18 &           &  \checkmark   &           & retrained      & 28.6\%   & 24.0\%   & 48.6\%   & 47.1\% \\
         19 &           &              & \checkmark & retrained     &   21.6\%   & 23.6\% & 48.5\%   &  \textbf{62.1}\%   \\
         20 &           & \checkmark    & \checkmark & retrained     &  \textbf{14.0}\%   &  \textbf{13.2}\%   &  \textbf{57.4}\%   &  61.4\% \\
    \midrule
         21 & Procrustes oracle & \checkmark  &   \checkmark        & retrained      &  12.1\% & 8.7\%  &  98.8\%   &  98.0\% \\
         22 & Wasserstein-Procrustes oracle & \checkmark &  \checkmark       & retrained      & 13.1\% & 10.0\% &  99.0\%   &  98.4\% \\
    \bottomrule
    \end{tabular}
    \label{tab:results}
\end{table*}

\subsection{Metrics} 
We use two metrics to evaluate our attack on the different scenarios: Equal Error Rate (EER) and Top 1 speaker accuracy.
Both metrics are computed for Female and Male speakers sets separately~\cite{tomashenkoVoicePrivacy2020Challenge}.

For all experiments, the EER is computed by scoring the $x$-vectors of the reconstructed $\mathbf{Trials_{Anon}*}$ set against the ones from the $\mathbf{Enroll}$ set, using cosine similarity. 
The lower the EER, the closer the reconstructed $x$-vectors are from the $\mathbf{Enroll}$ set, meaning we can find a link between the set attacked ($\mathbf{Trials}$) and the set used for the attack ($\mathbf{Enroll}$).
A low EER would imply the capacity of the attacker to break the \textbf{unlinkability} aspect of the speaker anonymization system.

The Top 1 speaker accuracy is computed by comparing $\mathbf{Trials}$ against $\mathbf{Trials_{Anon}*}$.
For each $x$-vector of $\mathbf{Trials_{Anon}*}$, we look for the nearest neighbor $x$-vector from $\mathbf{Trials}$ (using euclidean distance). 
The Top 1 speaker accuracy is the proportion of $x$-vectors from $\mathbf{Trials_{Anon}*}$ for which the closest $x$-vector in $\mathbf{Trials}$ is from the same speaker.
A high Top 1 speaker accuracy means a high success in reconstructing $x$-vectors close their clear counterpart and should raise concerns regarding the \textbf{non-invertibility} property of speaker anonymization methods.

\subsection{Results}
This section presents the experimental results (summarised in Table \ref{tab:results}) for the scenarios detailed in Section \ref{sec:scenarios}.

We add the four scenarios presented in the voice privacy 2020 evaluation plan~\cite{tomashenkoVoicePrivacy2020Challenge} (see Section \ref{subsec:VPC_scenars}), for which the EER metric was recomputed using the same data but with a cosine scoring (lines 1 to 4 of Table \ref{tab:results}).
Only the Equal Error Rate is computed here because the attackers proposed in the VPC cannot inverse the speaker anonymization function.

Lines 5 to 12 explore the supervised scenario (Section \ref{subsubsec:scenar_supervised}). 
We can see that Procrustes gives the same attack performance as the VPC baseline attacks under the same hypothesis (similar EERs in lines 8 and 12 than 3 and 4).
Regardless of the attack, the variation where the $x$-vector extractor is retrained on anonymized data consistently outperforms the one trained on clear data.
Procrustes (line 12) achieves a Top 1 speaker accuracy of 59.8\% (resp. 60.0\%) for Female (resp. Male) speakers, meaning that almost six times out of 10, the anonymized speaker $x$-vectors can be re-identified.
This raises concerns about the \textit{non-invertibility} aspect of the anonymization system.
The best performances are achieved by estimating a gender-dependent rotation and using PCA to reduce the $x$-vector dimensions (lines 8 and 12).

Lines 13 to 20 explore the unsupervised scenario (Section \ref{subsubsec:scenar_unsupervised}).
We can see that for almost every case, Wasserstein-Procrustes gives slightly worse results than the Procrustes counterpart, as no labels are available in this scenario.
We underline that the difference is usually around a few percent, so the distribution of $x$-vectors before and after anonymization is probably quite similar. Similar enough to get close results to when labels are available.

Lines 21 and 22 give results associated with the oracle approach (Section \ref{subsubsec:scenar_oracle}). 
Procrustes oracle (line 21) gives the best results among the previous experiments: 12.1\% (resp. 8.7\%) EER for Female (resp. Male) speakers, and 98.8\% (resp. 98.0\%) for the Top 1 speaker accuracy.
As expected, the results are worse for Wasserstein-Procrustes oracle (line 22), with a 99.0\% (resp. 98.4\%) for Female (resp. Male) Top 1 speaker accuracy. Interestingly, in this scenario, with only access to both clear and anonymized $x$-vectors sets (no label information), the majority of $x$-vectors could be re-identified by the attacker.

\section{Conclusion}
\label{sec:conclusion}
This paper investigates various linkability and invertibility attacks on the speaker anonymization baseline of the VPC 2020.
Using the challenge evaluation dataset, we approximate the anonymization function as a rotation between $x$-vectors domains before and after anonymization, using embedding alignment methods.
Procrustes (resp. Wasserstein-Procrustes) is used to estimate the rotation in a supervised (resp. unsupervised) way.
To improve performances, the attacker can compute the $x$-vectors thanks to an extractor retrained on anonymized utterances, estimate gender-dependent rotations, and apply PCA on the $x$-vectors.

Procrustes-based approaches are able to recover a large part of the mapping between clear and anonymized data; this leads to an EER which is lower than the EER calculated with the VoicePrivacy evaluation framework (line 4 and 12 of Table \ref{tab:results}).
It is also the case for the unsupervised scenario using Wasserstein-Procrustes, proving that label information is not mandatory to estimate accurate rotations.
Regarding the invertibility attack, 60\% Top 1 speaker accuracy is achieved in both scenarios, meaning that the inverse rotation can re-identify the majority of $x$-vectors. 
Oracle Procrustes experiment gives the upper bound for rotation approximation on the VPC baseline system: for Female and Male speakers, it could go down to 12.1\% and 8.7\% EER, respectively, with full access to the attacked sets.

In the unsupervised oracle scenario (e.g., full access to the unlabelled attacked sets), Wasserstein-Procrustes achieves one-to-one speaker matching between clear and anonymized counterparts with 99.0\% (resp. 98.4\%) accuracy for the Female (resp. Male) speakers.

The EER obtained shows that the \textbf{unlinkability} of a speaker anonymization system can be broken in the $x$-vector domain by an attacker using a rotation.
The Top 1 speaker accuracy leads to similar conclusions about the \textbf{non-invertibility}. It is of particular concern to notice that without any label information, an attacker with full access to clear and anonymized counterparts would be able to re-identify the majority of anonymized data.

Finally, these results show that there is room for improvement in current speaker anonymization systems. The unsupervised (Wasserstein-Procrustes) attack scenario seems to be an interesting approach to evaluate future anonymization methods' robustness against re-identification attacks.



\bibliographystyle{IEEEbib}
\bibliography{mybib}

\end{document}